\documentclass[conference]{IEEEtran}
\IEEEoverridecommandlockouts

\usepackage{amsmath,amssymb,amsthm}
\usepackage{stfloats}
\usepackage{graphicx}
\graphicspath{{figs/}}
\usepackage{booktabs}
\usepackage{dcolumn}
\usepackage{algorithm}
\usepackage{algorithmic}
\usepackage{tikz}
\usepackage{float}
\usetikzlibrary{shapes.geometric}
\usepackage{url}

\newtheorem{theorem}{Theorem}[section]

\newtheorem{definition}[theorem]{Definition}

\definecolor{algnotecol}{HTML}{50617A}
\newcommand{\algnote}[1]{{\color{algnotecol}\itshape // #1}}

\begin{document}

\title{AWE: Adaptive Weight Encoding for Exact Integer Matrix Products with
Fewer GEMMs on FP4 Tensor Cores}

\makeatletter
\newcommand{\linebreakand}{%
  \end{@IEEEauthorhalign}
  \hfill\mbox{}\par
  \mbox{}\hfill\begin{@IEEEauthorhalign}
}
\makeatother

\author{
\IEEEauthorblockN{Shun-ichiro Hayashi}
\IEEEauthorblockA{Graduate School of Informatics\\
Nagoya University\\
Aichi, Japan\\
hayashi@hpc.itc.nagoya-u.ac.jp}
\and
\IEEEauthorblockN{Daichi Mukunoki}
\IEEEauthorblockA{Information Technology Center\\
Nagoya University\\
Aichi, Japan\\
mukunoki@cc.nagoya-u.ac.jp}
\linebreakand
\IEEEauthorblockN{Tetsuya Hoshino}
\IEEEauthorblockA{Information Technology Center\\
Nagoya University\\
Aichi, Japan\\
hoshino@cc.nagoya-u.ac.jp}
\and
\IEEEauthorblockN{Takahiro Katagiri}
\IEEEauthorblockA{Information Technology Center\\
Nagoya University\\
Aichi, Japan\\
katagiri@cc.nagoya-u.ac.jp}
}

\maketitle

\begin{abstract}
Emulation of high-accuracy floating-point matrix multiplication, as in the
Ozaki scheme, splits the inputs into low-precision components and
multiplies them pairwise. These products must be error-free, and each is
an integer matrix product times a scale factor. FP4 Tensor Cores are the
fastest on the NVIDIA B200 and B300 but cannot hold INT8 operands.
The FP4 values scaled by 2 form the set
$S = \{0, \pm1, \pm2, \pm3, \pm4, \pm6, \pm8, \pm12\}$, which contains every
residue modulo 13, so with carries any integer splits into base-13 digits
that FP4 can store. Prior work splits each INT8 operand into 3 such digits
(limbs) with weights $(1, 13, 169)$ and multiplies them pairwise, 9 FP4
matrix multiplications (GEMMs) for INT8$\times$INT8. The classical ways to
reduce products, such as the Karatsuba and Toom--Cook methods, do not apply
as they stand: sums of limbs reach $\pm 24$ and leave $S$. This paper
asks how many FP4 GEMMs are needed for one integer matrix product. We
propose Adaptive Weight Encoding (AWE): the limbs take freely chosen
integer weights, the stored planes are linear combinations of limbs, and
the exact product is the sum of the FP4 GEMMs scaled by reconstruction
coefficients. For
each input range, we searched these choices for encodings with fewer
products and found INT8$\times$INT8 in 6 products and
INT4$\times$INT8 in 4. The formulation also holds modulo $m$, which
covers the residue number systems of Ozaki scheme II: for the FP64
significand, the 75 products of prior work are reduced to 59. The boundary in
product count between encodings with and without residues lies near input
width 15. We release the encodings found.
\end{abstract}

\section{Introduction}\label{sec:intro}
On recent GPUs the general-purpose FP64 units, which lie outside the tensor
cores, are being reduced, and FP64 throughput is a small fraction of the
low-precision throughput. High-accuracy matrix products on such machines are
obtained by emulation, exemplified by the Ozaki scheme~\cite{ozaki2012}: each
input matrix is split into several low-precision components, the pairwise
component products are computed, and the results are summed. The method
requires that the component products be computed without rounding error;
we call a matrix product computed with no rounding at any step an
\emph{error-free}, or exact, matrix product. Up to a scaling, the component products
are integer matrix products. An error-free integer matrix product is thus
not an end in itself: it is needed as the core operation of high-accuracy
floating-point emulation.

The first candidate for computing an integer matrix product without
rounding is the INT8 matrix unit, the integer Tensor Core:
products accumulate in INT32, so no rounding occurs as long as the
accumulator does not overflow. ozIMMU~\cite{ootomo2024} is the Ozaki scheme
implemented on this INT8 Tensor Core. Its INT8 instruction, however, is
being removed from the fifth-generation Tensor Core: on the data-center
B300, the fifth-generation Tensor Core
instructions have no INT8 variant, which the B200 has~\cite{cutlassbw,chen2608}. On such
GPUs, computing the integer matrix products on a floating-point tensor
core is an alternative. The Ozaki scheme has been implemented on FP8 tensor
cores~\cite{mukunoki2508,uchino}; since the 3-bit significand of FP8 cannot
represent an INT8 operand, each integer matrix product is then computed from
several FP8 matrix products. FP4 is present on both generations and is the
fastest format the Tensor Cores provide. Its element format E2M1 is
standardized in the OCP Microscaling (MX) specification~\cite{ocpmx} and
offered by several vendors. Against this background, Hayashi et al.~\cite{int53}
showed that the values representable in E2M1 can represent base-13 digits, and
gave a way to run the integer matrix products themselves on FP4 tensor
cores. Each INT8 operand is split into 3 base-13 digits and all
$3 \times 3$ pairs of digits are multiplied, so an INT8$\times$INT8
product requires 9 matrix products on the FP4 tensor core. In what follows,
these digits are called limbs, and a matrix product computed on the FP4
tensor core with FP32 accumulation is called an FP4 matrix product.

When the integer matrix products are computed as FP4 matrix products, the
total number of FP4 matrix
products is the number of integer matrix products the Ozaki scheme needs,
multiplied by the number of FP4 matrix products per integer matrix product.
The number of integer matrix products is set by the emulation scheme and
has long been studied. This paper asks how many FP4 matrix products one
error-free integer matrix product needs.

Let $S$ be the set of integer values
the matrix unit can store, let $(M_A, M_B)$ be the pair of integer widths,
and write $r$ for the number of matrix products, with every value drawn
from $S$, from which the error-free integer matrix product is reassembled.
Emulating an integer Tensor Core reduces to finding, for each $S$ and pair
of widths, constructions with few products. The known count is that of
the positional decomposition, $r = L_A L_B$, where $L$ is the number of
base-13 limbs each width needs. On FP4 no smaller count has been reported
at any width.

Three base-13 limbs represent every integer up to $\pm 732$, whereas INT8
needs only $\pm 127$. Adaptive Weight Encoding (AWE), the
proposed method, uses this headroom to reduce the number of products. It replaces the
positional weights $13^k$ by arbitrary integer weights, stores linear
combinations of limbs as planes, and treats the weights, the planes, and
the reconstruction coefficients as the design variables of a single
identity. The classical
methods derived from Karatsuba~\cite{karatsuba}, which reduce the number
of products of polynomial multiplication by the choice of evaluation points,
cannot be used as they stand on FP4, because sums of limbs fall into the
holes of the storable set. The reductions of this paper use planes of
a different form. The search declares a candidate
set of planes, enumerates every combination of those planes and the weights,
and verifies every solution independently. The only unit required is the one that computes FP4 matrix products: neither an integer tensor
core nor a wider floating-point unit is needed.

This paper makes four contributions.
\begin{enumerate}
 \item Section~\ref{sec:general} formulates the reduction of the product
       count in a general form: linear combinations of limbs under
       arbitrary integer weights, with reconstruction coefficients.
 \item Section~\ref{sec:catalog} gives, on the stored values of E2M1, a
       catalogue of encodings with fewer products than the positional
       decomposition. INT8$\times$INT8 takes 6 products instead of 9.
 \item Section~\ref{sec:boundary} relaxes the identity to a congruence
       modulo $m$ and applies it to the residue number systems of Ozaki
       scheme II. The moduli 11 and 13 also need only one product, and the FP64
       significand takes 59 products instead of 75 at reduction length
       $16{,}384$.
 \item Section~\ref{sec:release} releases the catalogue, with the search
       range declared on every record, and the verification scripts.
       Section~\ref{sec:boundary} compares the catalogue against residue
       number systems: for equal widths on the two sides, the input width
       above which residues need fewer products moves from 11 to 15.
\end{enumerate}

Section~\ref{sec:related} reviews related work and Section~\ref{sec:prep}
the preliminaries. Section~\ref{sec:method} presents the method and
Section~\ref{sec:catalog} the catalogue of solutions.
Section~\ref{sec:discussion} discusses the implications for
implementations.

\section{Related work}\label{sec:related}
\begin{figure*}[t]
\centering
\includegraphics[width=\textwidth]{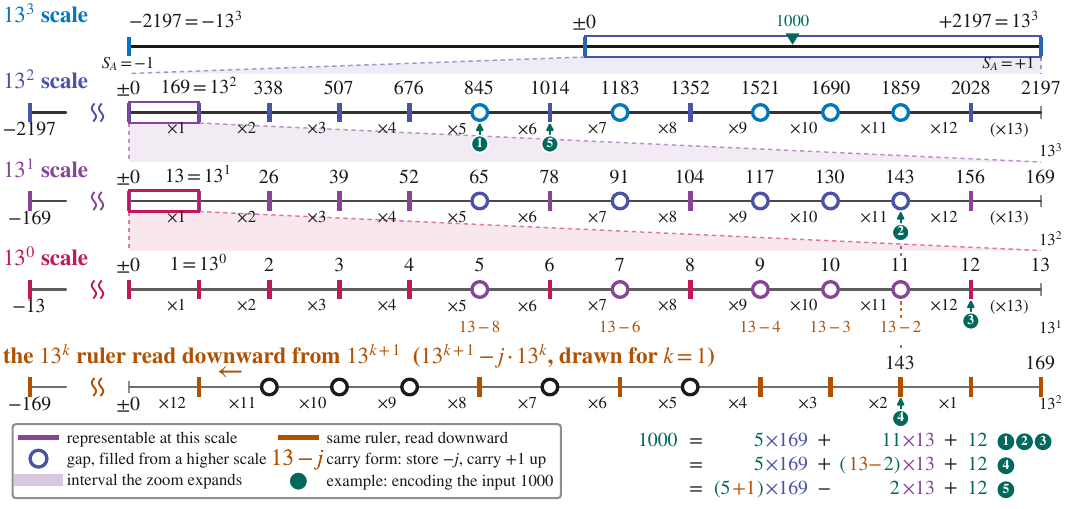}
\caption{The storage constraint imposed by base-13 positional notation, and
the decomposition procedure by carries (worked example: 1000). Each row is
the $13^k$ scale; open circles are values absent from $S$ (gaps). The
colour of a circle's ring indicates the higher scale that fills the gap;
the bottom ruler, read downward (carry form $13 - j$ with $j \in S$), fills
the remaining ones (legend in the figure).}
\label{fig:zoom}
\end{figure*}
\subsection{Exact products from units that round}
The Ozaki scheme is a family of methods that compute
floating-point matrix products to high accuracy by splitting the inputs into
low-precision components whose pairwise matrix products must be exact; up to
scaling, those component products are the integer matrix products this paper
studies. Ozaki scheme I~\cite{ozaki2012} is an error-free transformation that
splits a matrix product exactly into several matrix products, its count
growing as the product of the numbers of components. The framework has since been
implemented on other units. Mukunoki et al.~\cite{mukunoki2020} were
the first to run it on the low-precision units built for AI, obtaining DGEMM
from FP16 Tensor Cores, the first use of Tensor Cores for DGEMM emulation;
ozIMMU~\cite{ootomo2024} then implemented it on INT8
matrix units, and FP8 Tensor Cores followed~\cite{mukunoki2508}. The
second-generation Ozaki scheme II~\cite{ozakiII} computes the product in
parallel as residues modulo coprime moduli and reconstructs by the Chinese
remainder theorem; its FP8 realization~\cite{uchino} applies Karatsuba to a
2-component base-16 representation of each residue, reducing 4 component
products to 3. Hayashi et al.~\cite{int53} applied it to FP4
by showing that the E2M1 values can represent base-13 digits.

In these methods the representation is fixed first and the number of
products follows from it: a positional split into $L$ components requires
$L_A L_B$ products, a residue split requires one integer product per modulus. The count
follows from the representation and is not minimized over the choice of representation, and
the only reduction applied to it, in~\cite{uchino}, is a classical Karatsuba
step available because base-16 components fit an FP8 significand. What this
paper varies is the representation of the integer operands, not the
emulation scheme around it. The reduction in the number of products
therefore applies wherever an integer product has to be split into limbs
over a storable set, shown here for FP4,
and it is used inside Ozaki scheme I or II, as the layer that computes one
integer matrix product, rather than in place of them.

This paper builds on~\cite{int53} and differs from it in the following
respects. That work establishes the representation, uses it to implement
Ozaki schemes I and II, and remarks that the same
principle emulates an INT8 Tensor Core bit-exactly on FP4. Its 9 products for
INT8$\times$INT8 are the direct consequence of base-13 positional notation
and are not claimed to be optimal. Nor is the emulation cost examined at
other widths or with the limbs stored in a format other than FP4, such as
FP8. This paper takes that remark as its
subject, does not restrict the weights to the positional values, and replaces the
single count by the counts found by search for each pair of widths.

\subsection{Product counts without a storage constraint}
Karatsuba--Ofman~\cite{karatsuba} and Toom--Cook~\cite{toom,cook} reduce the
number of products of polynomial multiplication by the choice of evaluation
points. Minimizing the number of products has also been studied in algebraic
complexity theory~\cite{winograd77,fiduccia,bcs}, where the coefficients may
be drawn freely from a field. Our $r$ differs in that it is defined under the
constraint that every stored value lie in a finite set with holes.
Without the constraint, the product of
the two inputs is itself a single product, so a lower bound derived without
the constraint is not a lower bound on $r$; conversely, classical constructions become
unavailable format by format once their evaluation-point values or
intermediate sums leave the contiguous range (Section~\ref{sec:issues}). That
sums must remain representable was already noted by~\cite{uchino}; we state
this condition as thresholds on the significand width.

\subsection{Computer search for product decompositions}
Decompositions that reduce the number of products have also been sought by
computer search: by training a model that guides the search, as in
AlphaTensor~\cite{alphatensor}, and by giving the coefficient equations to
a SAT or constraint-programming solver~\cite{heule2019,deza2023}, the
latter of which can also prove that a declared space contains no solution.
These searches restrict
the combination coefficients to a small discrete set, a restriction chosen
to make the search finite. The constraint of this paper is not chosen: the
format imposes it on the values actually stored, and the set of those
values has holes.

\section{Preliminaries}\label{sec:prep}
This section restates the representation of Hayashi et al.~\cite{int53} in
the notation of this paper: the integer set stored by E2M1, and the base-13
limb decomposition with carries. The exact-accumulation condition also comes
from that work; Section~\ref{sec:kstar} states it in closed form.

\subsection{The FP4 format and the integers it stores}\label{sec:lattice}
Doubling the values representable in FP4 (E2M1) gives the integer set
$S = \{0, \pm1, \pm2, \pm3, \pm4, \pm6, \pm8, \pm12\}$ (15 values). $S$
contains the consecutive integers $[-4, 4]$ around the origin and is sparse
beyond them; the missing values are the \emph{holes} of $S$. A
floating-point format with an $m$-bit significand (hidden bit excluded)
represents consecutive integers up to $2^{m+1}$ (4 for E2M1 with $m = 1$; 16 for E4M3
with $m = 3$), and this quantity bounds the usable base from above (the
remaining holes in the stored values are avoided by carries,
Section~\ref{sec:limb}). The factor of two is absorbed into a block scale or
into the final shift of the reconstruction, and from here on we always count
in these doubled integers. A block scale is one scale factor applied to each
block of a fixed number of elements, treated in detail in
Section~\ref{sec:scalecond}. The stored bit patterns are plain E2M1, and no
run-time operation is added. A power-of-two factor is an exponent shift in
floating point and does not change whether a rounding occurs, so all exactness
arguments (including $K^*$ of Section~\ref{sec:kstar}) may be carried out
entirely under this integer convention.

\subsection{Base-13 limb decomposition and positional reconstruction}
\label{sec:limb}
An integer $N$ is decomposed as $N = \sum_i w_i a_i$ with $a_i \in S$. The
$a_i$ are the \textbf{limbs} and the $w_i$ the \textbf{weights}. Since $S$
contains a complete residue system modulo 13, the positional choice
$(1, 13, 169, \dots)$ is the basic form, and carries avoid the holes of $S$.
Writing $L$ for the number of limbs, the half-width covered contiguously by
$L$ limbs is $X_{\mathrm{carry}}(L)$, with
$X_{\mathrm{carry}}(L) = 13\, X_{\mathrm{carry}}(L{-}1) + 4$ and
$X_{\mathrm{carry}}(1) = 4$. Figure~\ref{fig:zoom} shows the multi-scale
rulers of this representation, the storage constraint they impose (the gaps),
and the role of the carries.

\subsection{Two difficulties of the positional representation}
\label{sec:issues}
With the positional weights $(1, 13, 169, \dots)$ fixed, two difficulties
appear in reducing the number of products.

First, the methods of Karatsuba and Toom--Cook~\cite{karatsuba,toom,cook}
cannot be used as they stand. They save products by computing one product
of sums, such as $(a_0 + a_1)(b_0 + b_1)$, and the sums must be storable.
The contiguous part of $S$ is $[-4, 4]$, and for $a_0, a_1 \in S$ the sum
$a_0 + a_1$ takes the values $5, 7, 9, 10, 11$ and values of 13 and above,
all of which are holes of $S$. Keeping the sums inside $S$ means restricting
the inputs, and the contiguous range of 2 limbs shrinks from 56, and even
with the widest choice of weight, 6 in place of 13, it is only 24.
Counted for a format with an $m$-bit significand, the sums fit the
contiguous part $2^{m+1}$ only if $m \ge 3$ for a Karatsuba-type step and
$m \ge 5$ for Toom-3. E2M1, with $m = 1$, is below both thresholds.

Second, the positional representation does not use the full range of its top limb.
Representing integer width $M$ takes the least $L$ with
$X_{\mathrm{carry}}(L) \ge 2^{M-1} - 1$, and since $13^L$ and $2^M$ rarely
match, the top limb has headroom beyond the target. For INT8 the 3
limbs cover $X_{\mathrm{carry}}(3) = 732$ where 127 is needed, more than 5
times the requirement. The headroom grows as the contiguous half-width
lies farther from a power of two (Appendix~\ref{app:ladder13}).

Adaptive Weight Encoding (AWE) uses this headroom to reduce the number of
products. It frees the weights
from the positional values, uses only the limb tuples whose sum planes stay
in $S$, and chooses the weights so that those tuples still cover the target
range contiguously; the products are reduced at the same number of limbs.

\subsection{The condition for exact FP32 accumulation}\label{sec:kstar}
The Tensor Core multiply--accumulate accumulates in FP32 (significand
$F = 24$ bits). Let $c_A, c_B$ bound the absolute values entering a product
(in the vocabulary of Section~\ref{sec:method}, the plane values, i.e.\
linear
combinations of limbs), and let $n_g$ be the number of products sharing one
accumulator ($n_g$ exceeds 1 when products with equal reconstruction
coefficients are merged into one accumulator,
Section~\ref{sec:runtime}). As long as the reduction
length $K$ satisfies
\[
K \le K^* = \left\lfloor \frac{2^F}{n_g\, c_A c_B} \right\rfloor,
\]
the accumulation is exact, with no rounding, for every input. For
$K > K^*$ the guarantee over all inputs is lost.

\section{Method}\label{sec:method}
\begin{figure*}[t]
\centering
\includegraphics[width=\textwidth]{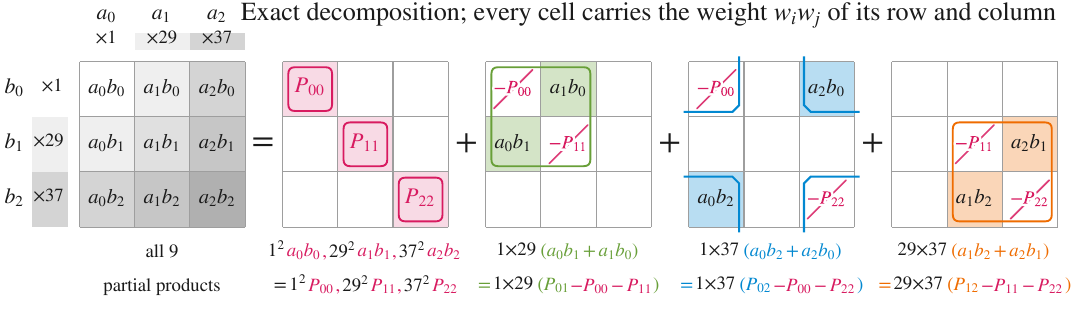}
\caption{Covering the 9 terms $a_i b_j$ with 6 products
(Section~\ref{sec:sixprod}). Outlines mark the terms in each product
(four corners for $P_{02}$), filled squares the terms that remain, slashed
squares the diagonal terms cancelled against $-P_{ii}$. The number under each
panel is its scale $w_i w_j$; the shading of the leftmost lattice
ranks the scales.}
\label{fig:coverage}
\end{figure*}

\subsection{Overview}
AWE has four components: (1)~a generalized representation that
chooses the number of limbs and the weights to match the integer width;
(2)~reduction of the product count through planes, the linear combinations
of limbs; (3)~encoding of weights into hardware block scales; and (4)~the
search procedure and its decision procedure.

\subsection{Generalized representation: integer width, limb count, free
weights}\label{sec:general}
The pair of limb counts, $p$ on the $A$ side and $q$ on the $B$ side,
is called the \textbf{cell} $(p, q)$.
\begin{definition}[Scheme]\label{def:scheme}
A scheme on the cell $(p, q)$ is a tuple of planes
$\alpha_k \in \mathbb{Z}^p$, $\beta_k \in \mathbb{Z}^q$, reconstruction
coefficients $\lambda_k$ ($k = 1, \dots, r$), and weights
$w \in \mathbb{Z}^p$, $v \in \mathbb{Z}^q$. The \textbf{admissible
tuples} are the limb vectors for which every plane value lies in $S$:
\[
T_A = \{a \in S^p : \alpha_k \cdot a \in S \ (k = 1, \dots, r)\}
\]
and $T_B$ likewise on the $B$ side. The scheme is required to satisfy, for
all $a \in T_A$ and $b \in T_B$, the identity
\[
\sum_{k=1}^{r} \lambda_k\, (\alpha_k \cdot a)(\beta_k \cdot b)
= (w \cdot a)(v \cdot b).
\]
We call $r$ the number of products. The reconstruction coefficients are
rationals $\lambda_k = \ell_k / D$ ($\ell_k \in \mathbb{Z}$) whose common
denominator $D$ is restricted to a power of two, implemented exactly by a
final shift.
\end{definition}
The half-width $X$ of the range $[-X, X]$ representable contiguously by
$w \cdot a$ ($a \in T_A$) is the \textbf{coverage} under $w$. Representing an
input of integer width $M$ requires $X \ge 2^{M-1} - 1$. Integer width $M$ in
this paper always refers to the symmetric range
$[-(2^{M-1}{-}1),\, 2^{M-1}{-}1]$; the two's-complement endpoint $-2^{M-1}$
is not handled. This is the same convention as in symmetric quantization. Products
of the significands of a floating-point format whose significand has $t$
bits, hidden bit included, are treated as integer products of width
$M = t + 1$ (9 for BF16, 12 for FP16, 25 for FP32, 54 for FP64). The
formulation places the $A$ and $B$ sides independently, so different
encodings for $A$ and $B$ are included from the start. Search and
verification use the coefficient matrix equation
$\sum_k \lambda_k \alpha_k \beta_k^{\mathsf T} = w v^{\mathsf T}$, which
implies the identity above. Every solution found is also checked over all
admissible pairs (Section~\ref{sec:verify}).

\subsection{Reducing the product count: combination planes with at most
2 terms}\label{sec:sixprod}
For INT8$\times$INT8 on the cell $(3,3)$ with target coverage 127, the common weights
$w = v = (1, 29, 37)$ and 6 planes per side turn the 9 products of the
direct decomposition, one product per pair of limbs, into 6. The identity holds for the products of encoded values
$(w \cdot a)(v \cdot b)$ and applies to the matrix product by summing along
the reduction dimension:
\[
P_{ii} = a_i b_i, \quad P_{ij} = (a_i + a_j)(b_i + b_j),
\]
\[
AB = -65 P_{00} - 261 P_{11} + 259 P_{22} + 29 P_{01} + 37 P_{02}
+ 1073 P_{12}.
\]
Figure~\ref{fig:coverage} shows how the 6 products cover the 9 terms
$a_i b_j$. Admissibility, the requirement that the combination-plane values
$a_i + a_j$ lie in $S$, constrains the weights, and the
coverage--product-count trade-off arises from this constraint. Whereas the direct
decomposition reuses 3 planes per side across 9 products, the 6
products use 6 planes per side one-to-one; at equal product count, the
number of plane loads can differ by up to a factor of $\sqrt{r}$. The
6-product solution is not unique: the same candidate
set contains a 6-product solution with weights $(1, 11, 39)$ and coverage
134. Since $11 = 13 - 2$ and $39 = 3 \times 13$, the construction remains
valid with weights close to the positional base 13, and its reduction-length
bound equals the main solution's. Neither weight vector is a geometric
sequence, so neither construction can be written in the form of
Karatsuba--Toom, whose weights are powers of the base. The full set of
solutions is collected in the catalogue (Section~\ref{sec:catalog}) and the
public repository; coefficients are in Appendix~\ref{app:tables}.

\subsection{Run-time procedure}\label{sec:runtime}
Algorithm~\ref{alg:runtime} shows the run-time procedure. The configuration,
that is, the weights, planes, and reconstruction coefficients, is a constant
fixed by the offline search, and no search runs at execution time. Encoding is a table lookup
per value. The values of combination planes are small integers, so the
additions that produce them are exact in integer or FP32 arithmetic alike.
The absolute value of the reconstruction's intermediate sum is at most
$\sum_k |\ell_k|\, c_A c_B K$, which fixes the bit width it needs; when it
exceeds 64 bits, the reduction is split into segments or wider integer
arithmetic is used.

\begin{algorithm}[tb]
\caption{Run-time procedure}
\label{alg:runtime}
\begin{algorithmic}
\REQUIRE configuration record: decomposition tables for the weights
$w, v$; planes $\{\alpha_k\}, \{\beta_k\}$; reconstruction $\{\ell_k\}$,
$D$; bound $K^*$. Integer matrices $A, B$ with reduction length $K \le K^*$
\ENSURE $C = AB$ exactly
\FOR{each limb $i$ of $A$ and of $B$}
  \STATE decompose each entry into $a_i$ by table lookup;
  \\ store the limb as E2M1 elements
\ENDFOR
\FOR{each combination plane of $A$ and of $B$}
  \STATE form its value by adding stored limbs
\ENDFOR
\FOR{each product $k = 1, \dots, r$}
  \STATE $P_k \leftarrow \text{FP4-GEMM}(\text{plane of } \alpha_k,\
  \text{plane of } \beta_k)$
  \\ \algnote{equal $\ell_k$ may share one accumulator
  ($n_g$, Section~\ref{sec:kstar})}
\ENDFOR
\STATE $C \leftarrow \bigl(\sum_k \ell_k P_k\bigr) / D$ in integer arithmetic of sufficient width
\\ \algnote{$D$ is a power of two; the division is one exact shift}
\RETURN $C$
\end{algorithmic}
\end{algorithm}

\subsection{Encoding weights in block scales}\label{sec:scalecond}
GPU FP4 matrix products apply a per-block scale to the E2M1 elements.
The OCP standard MXFP4 uses one E8M0 scale per 32 elements~\cite{ocpmx}, and the
NVIDIA extension NVFP4 one E4M3 scale per 16 elements~\cite{cutlassbw}. If a limb weight can be
written into the scale field, the multiplication outside the product
is no longer needed. The admissible weights differ by format: E8M0 represents only powers
of two, while E4M3 represents integers representable with a 3-bit significand
($13 = 1.625 \times 2^3$ is representable). The 6-product weights $(1, 29, 37)$
have odd parts exceeding 15 and fit neither format's scale, so the weight
multiplication is absorbed into the integer reconstruction. When the scale enters
the product from both sides, $A$ and $B$, no rounding occurs as
long as each factor's odd part is at most 15 and thus fits the E4M3
significand; the condition is determined by the set of storable values alone.

\subsection{Search procedure and decision procedure}\label{sec:search}
The search proceeds as follows. (1)~Declaration of the search space: the
cell, the candidate set of planes, and the weight range are stated
explicitly. (2)~Narrowing: the weight range is made finite by upper bounds
that follow from the target coverage alone. The derivation of the bounds is
kept in the records. Two further filters, that $\gcd(w) = 1$
whenever the coverage contains 1 and that coverage $X$ needs at least
$2X + 1$ distinct values $w \cdot a$, shrink the space by orders of
magnitude. (3)~Decision procedure: fixing the plane tuple and the weight $w$
of one side makes the coefficient equation linear in $(\lambda, v)$, so
existence and uniqueness are decided by exact rational arithmetic; this
avoids enumerating the other side's weights. Degenerate branches with
non-unique solution spaces are handled by enumerating integer points within the
weight bounds. The decision procedure is required to reproduce the known
solutions. (4)~Exhaustive execution over the
declared range.
Algorithm~\ref{alg:search} summarizes the procedure.

\begin{algorithm}[tb]
\caption{Offline derivation and verification of candidate solutions}
\label{alg:search}
\begin{algorithmic}
\REQUIRE integer widths $(M_A, M_B)$; candidate set of planes $\mathcal{F}$;
weight range $\mathcal{W}$ made finite by the weight bounds; number of
products $r$
\ENSURE list of configuration records that passed all-pairs verification
\FOR{plane tuples from $\mathcal{F}$ and weights $w$ from $\mathcal{W}$
passing the gcd and distinct-value-count filters}
  \STATE solve $\sum_k \lambda_k \alpha_k \beta_k^{\mathsf T} =
  w v^{\mathsf T}$ exactly for the reconstruction coefficients
  $\{\lambda_k\}$ and the weights $v$; take $D$ as the common denominator
  of the $\lambda_k$, required to be a power of two
  \IF{a solution exists}
    \STATE determine the coverages $X_A, X_B$ exhaustively
    \IF{the coverages meet the targets of $(M_A, M_B)$ and
    $AB = \sum_k \lambda_k P_k$ over all admissible input pairs}
      \STATE append one record:
      \\ \{weights, planes, $\lambda$, $D$, $K^*$, declared search range\}
    \ENDIF
  \ENDIF
\ENDFOR
\RETURN the records
\end{algorithmic}
\end{algorithm}

\subsection{Independent verification of solutions}\label{sec:verify}
Every solution is re-derived by a verifier written independently of the
search code: the reconstruction coefficients are solved again as exact
rationals, the coverage is recomputed, and the identity is checked over all
admissible input pairs. Exactness of the identity and storability of the
values it uses are separate claims, and both are checked. For every
construction reported here we confirm that every value the hardware holds
lies in the set its format can represent: the limbs, the plane values,
and the operands of a mixed-format product. The verification scripts, the
result hashes, and the environment are kept in the public repository.

\section{Catalogue of solutions and release}\label{sec:catalog}
The counts shown in this section, in the figures and in the table are
those of the encodings found by the searches so far. When a new encoding is found, the
catalogue in the public repository (Section~\ref{sec:release}) is updated
and the counts are updated with it.

\subsection{Solutions on the diagonal}
The counts found so far for equal widths $M_A = M_B = M$ are shown in
Figure~\ref{fig:toomk}; the coverages and full constructions are in
Appendix~\ref{app:tables} and the repository. The entries at $M \le 10$
come from sweeps over declared candidate sets of planes and weight ranges.
The entries at $M = 11$--$13$ and $15$--$17$ come from sweeps over
restricted weight families and plane patterns, and from lifting
(Section~\ref{sec:rect}); the weights and planes of each entry are in
Appendix~\ref{app:tables}. $M = 14$ stays at the 16 products of the direct
decomposition, and wider widths have not been searched. The figure also draws, for reference, the
direct decomposition $L^2$, the residue number system with the Chinese
remainder theorem at $K = 16{,}384$, and the Toom rank $2L - 1$ of
polynomial multiplication, which ignores the storage constraint.

\begin{figure*}[tb]
\centering
\includegraphics[width=\textwidth]{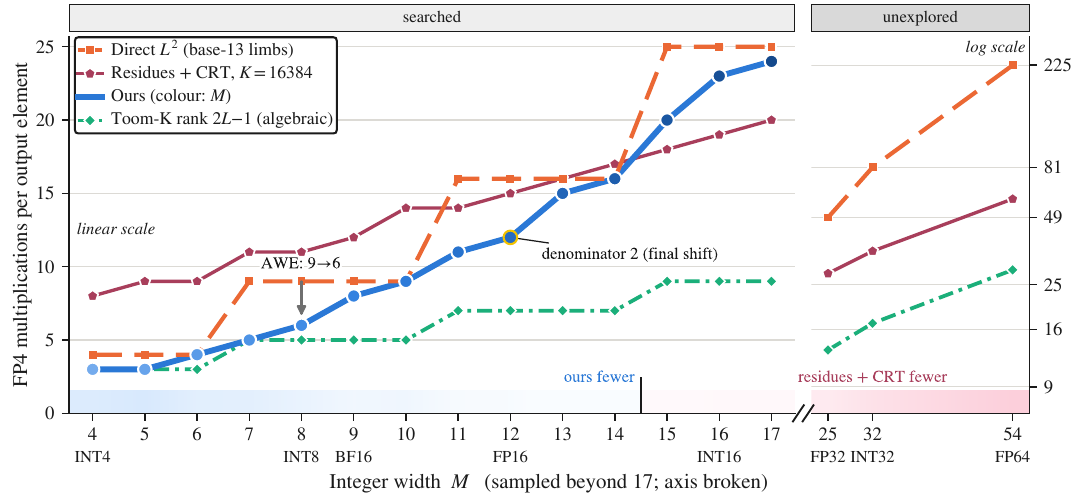}
\caption{Number of products at integer width $M$ ($M_A{=}M_B{=}M$),
linear scale on the left and logarithmic on the right. Blue: found by
search; the other series are references. The ribbon along the top marks
the widths searched. Format names under the axis: the integer width
$M = t + 1$ of each significand (Section~\ref{sec:general}).}
\label{fig:toomk}
\end{figure*}

\subsection{Asymmetric and rectangular solutions}\label{sec:rect}
INT4$\times$INT8 takes 4 products, 1 more than the 3 of
INT4$\times$INT4 and 2 fewer than the 6 of INT8$\times$INT8. The
catalogue of unequal width pairs, including that encoding, is kept in the public
repository. Figure~\ref{fig:planetri} shows the count for each pair of
widths $(M_A, M_B)$. The entries have four sources: restriction of the
diagonal entries, solutions found by search, monotonicity, and lifting.
Restriction sets to zero a limb that appears in no combination plane, which
leaves a scheme one limb narrower with the products that used that limb
removed. Monotonicity means that narrowing a width never raises the count.
Lifting appends one limb $b_q$ with weight $u$ on
the $B$ side of a solution on the cell $(p, q)$. The added part
$(w \cdot a)\, u\, b_q$ is reconstructed by the $p$ products of single
limbs $a_i b_q$, so $p$ more products give a solution on the cell
$(p, q{+}1)$, with the reconstruction denominator unchanged. The
purple frames in Figure~\ref{fig:planetri} mark entries obtained by this
rule alone; those cells themselves have not been searched. Where a cell
has both the purple frame and the yellow frame of a denominator 2, the
frame is split, yellow on the top and right and purple on the bottom and
left.

\begin{figure}[tb]
\centering
\includegraphics[width=\columnwidth]{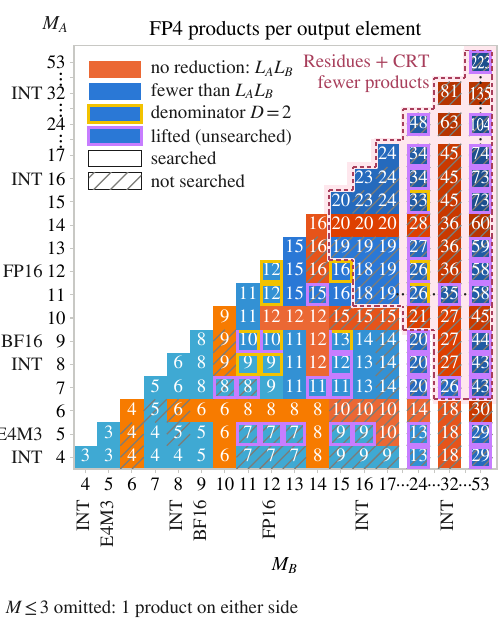}
\caption{Product counts found so far for the width pairs $(M_A, M_B)$.
Orange: at the direct count $L_A L_B$; blue: below it; darker: greater reduction.
Yellow frame: reconstruction denominator 2. Purple frame: obtained by
lifting, cell not searched. Hatched: not searched. Dotted staircase and
light red: where the residue system needs fewer products
(Section~\ref{sec:boundary}).}
\label{fig:planetri}
\end{figure}

\subsection{The boundary against residue number systems}\label{sec:boundary}
Relax the identity of Definition~\ref{def:scheme} from equality to a
congruence modulo $m$, restrict the reconstruction coefficients to integers,
and require additionally that the encodings on both sides represent every
residue. The result is a mod-$m$ scheme. Representing every residue means
that the map $a \mapsto w \cdot a \bmod m$ is surjective on $T_A$, and
likewise on the $B$ side. A residue number system splits the inputs into residues
modulo pairwise coprime moduli, computes the per-modulus products in
parallel, and reconstructs by the Chinese remainder theorem. Unique
reconstruction of the accumulated product requires the product of the
moduli to exceed $2 K B_A B_B$ (with $B = 2^{M-1}{-}1$ the input bound).
The same search applies to the per-modulus products: the moduli computed
by a single FP4 product are $2$ to $9$, $11$ and $13$, and 82 moduli are
computed by three products, listed in the public records. That $m = 11, 13$
are computed by a single product improves on the estimate of Hayashi et
al.~\cite{int53}, 1 product only up to modulus 9. On the block-scaled formats (MXFP4 and NVFP4) every
construction carries over as it stands, since the elements remain E2M1 and
the scale can be fixed at 1.
Choosing, by an integer program over this per-modulus table, the set of
moduli with the fewest products in total, and
comparing with the diagonal values found (Figure~\ref{fig:toomk}) gives
the boundary between the two approaches for the currently known
constructions. Against the direct decomposition
$L^2$, the residue system needs fewer products from integer width 11 on,
which is the comparison of the constructions of
Hayashi et al.~\cite{int53}. Comparing known constructions,
\textbf{the product-count reduction of this paper raises the break-even
width of the residue approach from 11 to 15} (the pentagon series of
Figure~\ref{fig:toomk}; reduction length $K = 16{,}384$). Only the residue
side depends on the reduction length $K$, which enters the required product
of the moduli. The comparison concerns the number of products; the run-time
boundary depends on the implementation of the reconstruction and is not
treated in this paper. The boundary is 15 for
$1{,}568 \le K \le 65{,}536$; at
shorter reduction lengths the boundary decreases, to 14 for
$1{,}352 \le K \le 1{,}567$ and 13 at $K = 1{,}024$. At width 15 the margin
is two products, 18 against our 20.
On the width plane the same comparison gives the staircase of
Figure~\ref{fig:planetri}; in most of the shaded cells the direct
decomposition is unreduced. At $(9, 24)$ both approaches need 20 products, and
at $(7, 24)$ and $(8, 24)$ the residue approach needs one product fewer
than a value obtained by lifting, on a cell not searched. These 3 cells
are left unshaded.

The choice of weights also determines, apart from the number of products,
which moduli a residue system can use. Two limbs of $S$ cover $\mathbb{Z}/m$ with the weight
fixed at the base only for $m \le 113$, together with $115$, $117$, $143$ and
$169$. Letting the weight be chosen per modulus makes $127$, $131$, $137$,
$139$, $149$, $151$ and $157$ usable, at weights $11$, $11$, $11$, $11$,
$34$, $35$ and $24$. At width 54, the FP64 significand, the construction of~\cite{int53}
uses 19 moduli and 75 products; on the wider set the integer program
selects 59, a factor of 1.3 below. That work notes that Karatsuba's method on the 7
moduli whose sum limbs are storable would bring its count to 68, but does
not use it in its evaluation. Table~\ref{tab:vsprior} compares these counts
with those of Hayashi et al.\ for three targets.

The moduli that free weights add enlarge the set available to the residue
side, but they do not change the boundary at widths 10--17: re-solving the choice of moduli with and
without them leaves every count at those widths unchanged, and they are
first chosen only near width 54.

\begin{table}[tb]
\centering
\caption{Product counts of the FP4 constructions of Hayashi et
al.~\cite{int53} and of this paper. The FP64 row compares residue number
systems at $K = 16{,}384$; the other rows compare encodings without
residues.}
\small
\begin{tabular}{@{}lrrr@{}}
\toprule
target & base-13~\cite{int53} & AWE & ratio \\
\midrule
INT8 $\times$ INT8 & 9  & \textbf{6}  & 1.5$\times$ \\
FP16 significand   & 16 & \textbf{12} & 1.3$\times$ \\
FP64 significand (CRT) & 75 & \textbf{59} & 1.3$\times$ \\
\bottomrule
\end{tabular}
\label{tab:vsprior}
\end{table}

\subsection{Incremental release and reproduction}\label{sec:release}
Each solution is stored as a JSONL record carrying a declaration of the
search range, the evidence, and the verification procedure. The declaration
states the cell, the target coverage, the candidate set and the weight
range, and the evidence consists of the script and result hashes and the
environment. All records and
the verification scripts are available at
\url{https://github.com/FP4-is-All-you-Need/BLAS}.

\section{Discussion}\label{sec:discussion}
\subsection{Implications for implementations}\label{sec:impl}
Whether the product-count reduction is beneficial in an implementation depends on
the cost of producing the planes, which differs greatly between writing
the produced planes back to Shared Memory and deriving them in registers.
$K^*$ is a segmentation
threshold, not a feasibility limit: it sets how often the accumulation must
be segmented, and summing the segments in sufficiently wide integers extends
the reduction length arbitrarily. Practical reduction lengths (a few thousand)
are well below the $K^*$ of every construction in this paper.

A machine that has only the unit for FP4 matrix products can still emulate
operations of the formats it lacks, at a constant factor. For instance, an
E4M3
value is an integer significand (absolute value at most 15) times a power of
two, and the significand fits in 2 limbs, so an exact FP8$\times$FP8
product can be replaced by the Karatsuba-type 3 products of the $(2,2)$
cell. This replacement applies only when the exponents can be aligned, that
is, when the inputs can be treated as fixed point. For general floating-point inputs with
per-element exponents, aligning the exponents within each block is an
additional step.

\subsection{Theoretical rate ratios against native units}\label{sec:ceilings}
The same estimate extends to other native formats. Table~\ref{tab:native}
shows the theoretical ratio to native throughput when this construction
computes the exact product of each format's significand. The values for all three
generations are based on published dense rates. Of the three compute
capabilities, sm\_120 corresponds to the RTX PRO 6000 Blackwell, sm\_100 to
B200, and sm\_103 to B300; sm\_120 and sm\_100 have the same throughput
ratio in every row. The values include neither plane production nor
reconstruction. The hardware unit of
each row is the unit on that generation that multiplies
in that format with one product, not a mixed-precision library: the format's own native unit, except that the fifth-generation Tensor Core
instructions of B300 have no integer variant~\cite{cutlassbw}, so its integer
rows are compared with one exact BF16 product.

The table is read as follows. The target operation is the Tensor Core
operation to be computed exactly. ``FP4 GEMMs'' gives the number of FP4
matrix products the construction needs. The hardware unit is the unit that
computes the target operation in one product, together with its published
peak rate. The AWE peak is the published FP4 Tensor Core peak rate divided
by the number of FP4 GEMMs. The throughput ratio is the AWE peak divided by
the hardware unit's peak rate. The hardware peak rates are published dense
rates of one GPU: the RTX PRO 6000 Blackwell Workstation Edition for
sm\_120~\cite{nvrtxpro}, and one GPU of HGX B200 and HGX B300 for sm\_100
and sm\_103~\cite{nvhgx}; the FP4 peak rates are 2.015, 9.0 and 13.5
PFLOPS. The AWE peak assumes that the FP4 Tensor Core
runs at its peak; it is not a measured value. A throughput ratio of
$1.00\times$ means that the exact product takes the same time as one
product on the hardware unit, and $0.33\times$ means three times as long.
In the INT8$\times$INT8 row, for instance, the FP4 peak of B200 is 9.0
PFLOPS and 6 products are needed, so the AWE peak is 1.50 POPS, which is
$0.33\times$ the 4.5 POPS of the INT8 unit. On B300 the fifth-generation
Tensor Core instructions have no INT8 variant, so the comparison is with
the BF16 unit: 13.5 PFLOPS divided by 6 gives 2.25 POPS, the same value as
the 2.25 PFLOPS of the BF16 unit, so the throughput ratio is $1.00\times$.

For the
INT8$\times$INT8 row, the BF16 substitute depends on the reduction
length $K$. Up to
$K^* = \lfloor 2^{24}/127^2 \rfloor = 1{,}040$ the single BF16 product is exact as it
stands; in practice this means blocks of $1{,}024$. Beyond it, the
reduction must be split into segments of at most $1{,}024$ and the partial
sums added exactly in a wider integer accumulator. Segmentation beyond
$K^*$ is needed in every construction, but the ratio of the two $K^*$ makes
it about 112 times as frequent for BF16 as in the 6-product construction.
The throughput ratios
exclude this cost ($\dagger$); how much of it an implementation overlaps
with other work is outside this paper. FP8 cannot replace the product with
a single GEMM at any $K$, because its contiguous integer range, $\pm 16$,
does not cover the INT8 input range; the 6-product construction needs no segmentation up to
$K^* = 116{,}508$. The narrower integer rows remain exact far beyond practical
lengths, as the bound scales as $2^{24}$ over the product of the two operand
ranges.

The significand ranges include the hidden bit ($\pm 255$ for BF16). The product of two BF16
significands takes 8 FP4 products; excluding plane production and
reconstruction, the resulting throughput is 0.50 times that of the native BF16 unit on
sm\_120 and B200 and 0.75 times on B300. The FP16 significand ($\pm 2{,}047$
with the hidden bit) takes the 12 products of the width-12 entry of
Figure~\ref{fig:toomk}, giving 0.33 times the throughput of the native FP16 unit on sm\_120 and
B200 and 0.50 times on B300; FP16 and BF16 share one published rate on all
three.
Native multiply--accumulate rounds, whereas
this construction is exact and bit-reproducible throughout, so, excluding plane
production and reconstruction, the arithmetic time relative to one native
product is the reciprocal of the throughput ratio.

The FP64 class, with a 53-bit significand and width 54, is omitted from
the table: as Section~\ref{sec:boundary} states, at that width the residue
approach needs fewer products, 59 at $K = 16{,}384$, so the direct
decomposition is no longer the preferable construction, and the residue
system's Chinese-remainder reconstruction is a cost this table excludes.

\begin{table*}[tb]
\centering
\caption{Theoretical peak rate of the target Tensor Core operation when AWE
replaces it by FP4 matrix products, and its throughput ratio to the peak
rate of the hardware unit that computes that operation in one product
(three Blackwell compute capabilities, one GPU, dense, published peak
rates~\cite{nvhgx,nvrtxpro}; not measurements). Rates in POPS, $10^{15}$
operations per second; for the floating-point units this is PFLOPS. $\dagger$: the BF16 substitute
is exact only to $K^* = 1{,}040$ (see text). $\ddagger$: reconstruction
denominator 2, one final shift.}
\footnotesize
\setlength{\tabcolsep}{2.4pt}
\newcolumntype{d}[1]{D{.}{.}{#1}}
\begin{tabular}{@{}lrl d{1.3} d{1.3} l l d{1.2} d{1.2} l l d{1.2} d{1.2} l@{}}
\toprule
 & & \multicolumn{4}{c}{sm\_120 (RTX PRO 6000)} & \multicolumn{4}{c}{sm\_100 (B200)} & \multicolumn{4}{c@{}}{sm\_103 (B300)} \\
\cmidrule(lr){3-6}\cmidrule(lr){7-10}\cmidrule(l){11-14}
target & FP4 & \multicolumn{2}{l}{hardware unit} & \multicolumn{1}{c}{AWE peak} & throughput & \multicolumn{2}{l}{hardware unit} & \multicolumn{1}{c}{AWE peak} & throughput & \multicolumn{2}{l}{hardware unit} & \multicolumn{1}{c}{AWE peak} & throughput \\
operation & GEMMs & type & \multicolumn{1}{c}{[POPS]} & \multicolumn{1}{c}{[POPS]} & ratio & type & \multicolumn{1}{c}{[POPS]} & \multicolumn{1}{c}{[POPS]} & ratio & type & \multicolumn{1}{c}{[POPS]} & \multicolumn{1}{c}{[POPS]} & ratio \\
\midrule
INT8 $\times$ INT8         & 6  & INT8 & 1.008 & 0.336 & 0.33$\times$ & INT8 & 4.5  & 1.50 & 0.33$\times$ & BF16 & 2.25 & 2.25 & \textbf{1.00}$\times^\dagger$ \\
INT4 $\times$ INT8         & 4  & INT8 & 1.008 & 0.504 & 0.50$\times$ & INT8 & 4.5  & 2.25 & 0.50$\times$ & BF16 & 2.25 & 3.38 & \textbf{1.50}$\times$ \\
$\{0, \pm 1\} \times$ INT8 & 3  & INT8 & 1.008 & 0.672 & 0.67$\times$ & INT8 & 4.5  & 3.00 & 0.67$\times$ & BF16 & 2.25 & 4.50 & \textbf{2.00}$\times$ \\
FP8 significand            & 3  & FP8  & 1.008 & 0.672 & 0.67$\times$ & FP8  & 4.5  & 3.00 & 0.67$\times$ & FP8  & 4.5  & 4.50 & 1.00$\times$ \\
BF16 significand           & 8  & BF16 & 0.504 & 0.252 & 0.50$\times$ & BF16 & 2.25 & 1.13 & 0.50$\times$ & BF16 & 2.25 & 1.69 & 0.75$\times$ \\
FP16 significand           & 12$^\ddagger$ & FP16 & 0.504 & 0.168 & 0.33$\times$ & FP16 & 2.25 & 0.75 & 0.33$\times$ & FP16 & 2.25 & 1.13 & 0.50$\times$ \\
\bottomrule
\end{tabular}
\label{tab:native}
\end{table*}

A narrower input value set allows the same theory to reduce the product count
further and raises $K^*$. Weights in $\{0, \pm 1\}$, as in the BitNet family of
quantized language models~\cite{bitnet}, fit in 1 limb, so
$\{0, \pm 1\}\times$INT8 takes 3 products, and the bound on the reduction
length is $K^* = 2{,}097{,}152$.

A lower product count does not by itself make this construction the one to
choose. An INT32
accumulator holds $\{0, \pm 1\} \times$ INT8 exactly, so on GPUs whose
fifth-generation Tensor Cores support INT8 one product suffices with no rounding, and
the FP4 construction provides no benefit. The construction is useful only on
GPUs without that instruction, such as B300. There
the exact alternative with the fewest products is one BF16 product: BF16 is the narrowest
floating-point format whose significand represents $\pm 127$ in a single limb, and for this
value set $K^* = \lfloor 2^{24}/127 \rfloor = 132{,}104$ keeps it exact at any practical
reduction length. One FP8 limb represents only
$\pm 15$, so FP8 needs 2 limbs, and FP4 needs 3. Which one is preferable on a
given generation, in this theoretical rate comparison, follows from the value set and the $K^*$
estimate, before any measurement.

\section{Conclusion}\label{sec:concl}
For exact integer matrix products over the values representable in FP4
(E2M1), we gave a general method that reduces the number of products through
free linear combinations of limbs with free weights, and a catalogue of the
encodings found by search for each pair of integer widths. The principal
entry is INT8$\times$INT8, where the 9 products of the direct decomposition
become 6. The same formulation also holds modulo $m$ and applies to
residue number systems: the moduli 11 and 13 also need only one product, and
for the FP64 significand, width 54, the 75 products of prior work are reduced to
59 at reduction length $16{,}384$. The catalogue is released in verifiable form, with the search range
declared on every record, and will be extended incrementally.

Two problems remain open. First, no 5-product encoding of INT8$\times$INT8 was
found with independent planes on the two sides and components of absolute
value up to 2, or with the same planes on both sides and components of
absolute value up to 3 and $\ell_1$ norm up to 5. Beyond these ranges the
question is unexplored. Second, whether
components $\{0, \pm1\}$ suffice for the planes remains a conjecture.

\section*{Acknowledgment}
This work was supported by JSPS KAKENHI Grant Number JP25K24387. Part of this
work was supported by the Joint Usage/Research Center for Interdisciplinary
Large-scale Information Infrastructures (JHPCN) in Japan (Project ID:
jh260065).

Generative AI assistants (Anthropic Claude and OpenAI Codex) were used under
the authors' direction to draft and translate portions of the manuscript text,
to write the figure-plotting scripts, and to assist in developing the search
and verification code. All methods, results, and claims were designed,
verified, and approved by the authors, who take full responsibility for the
content of this paper.

\appendices
\section{Coefficient tables}\label{app:tables}
\begin{figure*}[t]
\centering
\includegraphics[width=0.8\textwidth]{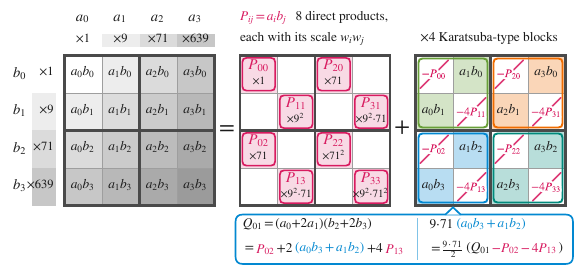}
\caption{The 16 terms $a_i b_j$ covered by 12 products ($M = 12$, the FP16
significand; limbs in increasing weight $1, 9, 71, 639$). Each bold
$2 \times 2$ block is one outer pair: a Karatsuba-type step of 2 direct products
$P_{ij}$ (pink) and 1 plane product $Q_{kl}$ (coloured). Fills: cells that
remain; strikes: cells cancelled.}
\label{fig:m12}
\end{figure*}
Coefficients of the principal constructions follow. The complete
coefficients are in the records (JSONL) of the public repository.

\medskip\noindent\textbf{INT8$\times$INT8 in 6 products.}
\[
A = a_0 + 29 a_1 + 37 a_2, \qquad B = b_0 + 29 b_1 + 37 b_2.
\]
Products $P_{ii} = a_i b_i$,
$P_{ij} = (a_i + a_j)(b_i + b_j)$;
\[
AB = -65 P_{00} - 261 P_{11} + 259 P_{22} + 29 P_{01} + 37 P_{02}
+ 1073 P_{12},
\]
$D = 1$. The reduction-length bound depends on the range of stored values
the encoding uses: capping every stored value at $\pm 6$ still attains
coverage 148, and gives $c_A = c_B = 6$ and
$K^* = \lfloor 2^{24}/6^2 \rfloor = 466{,}033$, whereas an encoder that
uses all of $S$ gives $c_A = c_B = 12$ and
$K^* = \lfloor 2^{24}/12^2 \rfloor = 116{,}508$. An alternative 6-product
solution has $w = v = (1, 11, 39)$ and coverage 134; its reconstruction
coefficients, in the order $(P_{00}, P_{11}, P_{22}, P_{01}, P_{02}, P_{12})$,
are $(-49, -319, 1053, 11, 39, 429)$.

\medskip\noindent\textbf{INT4$\times$INT8 in 4 products.}
\[
A = 3 a_0 + 5 a_1, \qquad B = 5 b_0 + 3 b_1 + 42 b_2.
\]
The values of the $k$-th planes are, in order,
\[
(\alpha_k \cdot a)_{k=1}^{4} = (a_1,\ a_0,\ a_0 + 2a_1,\ 2a_0 + a_1),
\]
\[
(\beta_k \cdot b)_{k=1}^{4} = (b_0,\ b_1 - b_2,\ b_2,\ b_0 + 2b_1),
\]
and with $P_k = (\alpha_k \cdot a)(\beta_k \cdot b)$
\[
2AB = 35 P_1 - 42 P_2 + 210 P_3 + 15 P_4
\]
($D = 2$). An encoding that keeps the plane values within $\pm 3$ on the
$A$ side and $\pm 8$ on the $B$ side has $c_A = 3$, $c_B = 8$ and
$K^* = 699{,}050$.

\begin{figure}[!t]
\centering
\includegraphics[width=\columnwidth]{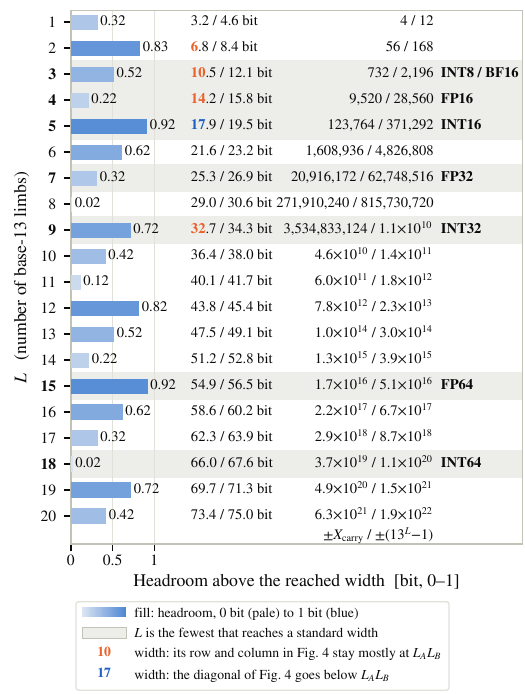}
\caption{The range of $L$ base-13 limbs: the contiguous coverage
$\pm X_{\mathrm{carry}}(L)$, the maximum $\pm(13^L{-}1)$, and the
headroom above the represented width [bit].}
\label{fig:widths13}
\end{figure}

\medskip\noindent\textbf{FP16$\times$FP16 in 12 products.}\quad
The weights $(1, 71, 9, 639)$ of the diagonal entry $M = 12$, sorted in
increasing order with the limbs renumbered:
\[
A = a_0 + 9 a_1 + 71 a_2 + 639 a_3, \qquad B = b_0 + 9 b_1 + 71 b_2 + 639 b_3.
\]
With $P_{ij} = a_i b_j$ and $Q_{kl} = (a_{2k} + 2a_{2k+1})(b_{2l} + 2b_{2l+1})$
for $k, l \in \{0, 1\}$,
\[
2AB = \sum_{k, l = 0}^{1} 71^{k+l} \left( -7 P_{2k,2l} + 126 P_{2k+1,2l+1} + 9 Q_{kl} \right),
\]
$D = 2$, $c_A = c_B = 12$, $K^* = 116{,}508$. Of the weights, 9 has odd part
at most 15 and fits an E4M3 block scale, whereas 71 and 639 do not and are
applied in the integer reconstruction (Section~\ref{sec:scalecond}). The inner
pairs $(a_0, a_1)$ and
$(a_2, a_3)$, of base 9, take one Karatsuba-type step with the planes
$a_0 + 2a_1$ and $a_2 + 2a_3$, and the four outer pairs, of base 71, are direct
products, $4 \times 3 = 12$. Figure~\ref{fig:m12} shows this structure, and
the form $U + sV$ of the diagonal weights describes the same construction by
its outer pairs. With the plane
coefficient $u = 2$, extracting the cross term leaves the denominator 2 in
the reconstruction coefficients, removed by the final shift of
Definition~\ref{def:scheme}. The choice $u = 2$ suits $S$ because the elements
of $S$ outside the contiguous part $[-4, 4]$, namely $\pm6, \pm8, \pm12$,
are all even, and half of any even element of $S$ is again in $S$. The
plane $a + 2b$ has an admissible $a$ for every $b \in S$, whereas $u = 3$
excludes $b = \pm12$ and $u = 4$ also $\pm8$.

\medskip\noindent\textbf{Diagonal weights for $M = 11$--$13$ and $15$--$17$.}\quad
Table~\ref{tab:diagweights} lists the weights of these entries.

\begin{table}[H]
\centering
\caption{Diagonal entries $M = 11$--$13$ and $15$--$17$ of
Figure~\ref{fig:planetri}: limb weights $w_i$, number of products $r$, and
coverage $\pm X$.}
\small
\setlength{\tabcolsep}{3pt}
\begin{tabular}{@{}lrrrrrrrl@{}}
\toprule
$M$ & $w_0$ & $w_1$ & $w_2$ & $w_3$ & $w_4$ & $r$ & $\pm X$ & note \\
\midrule
11 & 3 & 29 & 33 & 319   &         & 11 & 1{,}232  & \\
12 & 1 & 71 & 9  & 639   &         & 12 & 2{,}160  & $D = 2$ \\
13 & 4 & 11 & 91 & 1183  &         & 15 & 5{,}082  & \\
15 & 1 & 9  & 41 & 369   & 3987    & 20 & 16{,}828 & \\
16 & 1 & 7  & 61 & 689   & 8957    & 23 & 38{,}764 & \\
17 & 4 & 11 & 91 & 1183  & 15379   & 24 & 66{,}598 & lifted from $M = 13$ \\
\bottomrule
\end{tabular}
\label{tab:diagweights}
\end{table}

\noindent Three of these, $M = 11, 12, 15$, share one structure. Weights of the form $(p, q, sp, sq)$
split the value into $U + sV$ with $U = p a_0 + q a_1$ and
$V = p a_2 + q a_3$, so one Karatsuba-type step applies to the pair $(U, V)$
through the two planes $a_0 + u a_2$ and $a_1 + u a_3$, saving 4 products;
a further plane inside $U$ or $V$ saves a fifth. This gives $M = 12$ at
$(p, q, s, u) = (1, 71, 9, 2)$, $M = 11$ at $(3, 29, 11, 1)$ with the
extra plane $a_2 + a_3$, and $M = 15$ at $(1, 9, 41, 1)$ on 4 limbs
with a fifth limb appended and the extra plane $a_0 + a_1$. The $M = 16$
entry comes from a different family: a 4-limb solution with the two
planes $a_0 - a_1$ and $a_0 - a_2$, lifted by $z = 8{,}957$. In every
lifted entry the appended weight ($w_4$ in the table) was chosen by
exhaustive enumeration rather than by the positional value.

Besides the weights, each entry is fixed by its combination planes. The planes of the
$(p, q, sp, sq)$ family are given above, and $M = 13$ uses the single
plane $a_0 + a_1$. Full reconstruction coefficients for every entry are in
the records.

\medskip\noindent\textbf{INT8$\times$INT8 with 2 FP8 products and 1 FP4
product.}\quad The inputs of width 8 are encoded as $A = 13 h + 5\varepsilon$
and $B = 13 h' + 5\varepsilon'$, with $h, h'$ stored in FP8 and
$\varepsilon, \varepsilon' \in \{0, \pm1, \pm2, \pm3, \pm4, \pm6, \pm8\}$ in
FP4. The three products are $P_1 = h h'$ and
$P_2 = (h + \varepsilon)(h' + \varepsilon')$ in FP8 and
$P_3 = \varepsilon \varepsilon'$ in FP4, with
\[
AB = 104 P_1 + 65 P_2 - 40 P_3,
\]
$D = 1$. An encoding that keeps the plane values within $\pm 14$ has
$c_A = c_B = 14$ and $K^* = \lfloor 2^{24}/14^2 \rfloor = 85{,}598$. On
generations where FP8 runs at half the FP4
rate, this costs the equivalent of 5 FP4 products, fewer than the 6 of the
FP4-only scheme.

\section{The base-13 limb counts against binary widths}\label{app:ladder13}
Figure~\ref{fig:widths13} compares the contiguous coverage
$X_{\mathrm{carry}}(L) = (13^L - 1)/3$ of Section~\ref{sec:limb} with
the standard binary widths. It determines the least number of limbs an
integer width $M$ needs,
$L(M) = \min\{L : X_{\mathrm{carry}}(L) \ge 2^{M-1} - 1\}$, the direct
count $L(M)^2$, and the format marks of Figures~\ref{fig:toomk}
and~\ref{fig:planetri}. The headroom in the figure is
$\log_2(X_{\mathrm{carry}}(L) + 1) - (M_L - 1)$ bits, where $M_L$ is the
largest integer width that $L$ limbs represent; it is the room that the
free-weight constructions use. The largest magnitude $13^L - 1$ is
three times the contiguous half-width $X_{\mathrm{carry}}(L)$, an offset of
about $\log_2 3 = 1.585$ bits between the two bit counts. From row to row
the headroom increases by about the fractional part of $\log_2 13$,
$0.7004$, and wraps past 1. The
orange rows and columns of Figure~\ref{fig:planetri}, which stay mostly at the
direct count, lie at the widths 6, 10 and 14, the largest widths that 2, 3 and 4
limbs represent, shown in orange in Figure~\ref{fig:widths13} as well. At 17, the largest width
of 5 limbs, the diagonal cell goes below the direct count, shown in blue.

\end{document}